\documentclass{article}

\usepackage{PRIMEarxiv}

\usepackage[utf8]{inputenc} 
\usepackage[T1]{fontenc}    
\usepackage{hyperref}       
\usepackage{url}            
\usepackage{booktabs}       
\usepackage{amsfonts}       
\usepackage{nicefrac}       
\usepackage{microtype}      
\usepackage{lipsum}
\usepackage{fancyhdr}       
\usepackage{graphicx}       
\graphicspath{{media/}}     

\title{ Minimal warm inflation and TCC
\thanks{\textit{\underline{Citation}}: 
\textbf{Authors. Title. Pages.... DOI:000000/11111.}} 
}

\author{
  Vahid Kamali,  \\
  Department of Physics, Bu-Ali Sina (Avicenna) University, Hamedan 65178,
016016, Iran. \\
  \texttt{Email:vkamali@ipm.ir} \\
  {\bf Hossein Moshafi,}  \\
  School of Astronomy, Institute for Research in Fundamental Sciences (IPM), P. O. Box 19395-5531, Tehran, Iran \\
  \texttt{Email:moshafi@ipm.ir} \\
   \And
  Saeid Ebrahimi,  \\
  Department of Physics, Bu-Ali Sina (Avicenna) University, Hamedan 65178,
016016, Iran. \\
  \texttt{Email:saeid.ebrahimi1365@gmail.com} \\
}

\begin{document}
\maketitle

\begin{abstract}
The minimal warm inflation model was constructed as a warm inflation setup with direct interaction between inflaton and (non-Abelian) gauge fields. The model was shown to be compatible with observation for some forms of potential. As a result of direct analysis of CMB data, the model presents a reasonable phase-space of its parameter compatible with observation and the Trans-planckian censorship conjecture (TCC). 
\end{abstract}

\keywords{Warm inflation \and TCC }

\section{Introduction}
A cosmological theory called the Hot Big Bang (HBB) has been widely accepted as one that can explain observational data in a systematic way \cite{Planck:2018vyg,Planck:2019nip}. There are some inquiries in this context that are best explained by adding a short period of rapid expansion during the early evolution of the Universe. Accelerated expansion during early time is generally based on quantum field theory (QFT), mainly scalar fields \cite{Linde:1981mu,Guth:1980zm,Albrecht:1982wi,Sato:1980yn,Starobinsky:1979ty,Starobinsky:1980te}. 
Perturbation of inflaton as quanta of the scalar field that is responsible for accelerated expansion is the seed of large-scale structures (LSS) \cite{Mukhanov:1981xt,Hawking:1982cz,Guth:1982ec,Starobinsky:1982ee,Bardeen:1983qw}.
The CMB temperature map can also be explained by linear perturbation theory in a curved manifold (FLRW).
The standard inflation model employs a scalar field that rolls a nearly flat potential during the accelerated phase of expansion known as slow-roll epoch.
It is followed by the reheating epoch, during which the inflaton has transferred its energy, mainly to light particles \cite{Traschen:1990sw, Kofman:1997yn,Shtanov:1994ce,Kofman:1994rk,Allahverdi:2010xz}.
There is an alternative explanation for this two-part sector of early time evolution, in which the inflaton interacts with light fields in a thermal bath during slow-roll expansion \cite{Berera:1995ie}.
Within this so-called warm inflation scenario, the inflaton-dominated era connects smoothly to the radiation-dominated era without any separate reheating epoch.
Building a QFT model with a Direct coupling between inflaton and a limited number of light fields presents some challenges \cite{Yokoyama:1998ju}.
To resolve the inquiries, the warm little inflation model was presented, which fits with observations in a weak dissipative regime \cite{Bastero-Gil:2016qru}. The model has been extended by a scenario compatible with observations in a high dissipative regime  \cite{Bastero-Gil:2019gao}. In Ref. \cite{Kamali:2019ppi}, the direct coupling of inflatons and $SU (2)$ gauge fields, with the radiation equation of state $ P = \frac{\rho}{3} $, is investigated within a warm inflation context for the first time.
In this case, the dissipation parameter is proportional to the Hubble parameter, keeping it out of swampland and in line with observations \cite{Motaharfar:2018zyb,Kamali:2019wdh,Ebrahimi:2021see}.  
The idea of using Chern-Simons interaction to find a thermal bath during early time acceleration epoch was extended by introducing a minimal warm inflation model with a temperature-dependent dissipation term \cite{Berghaus:2019whh} and extended in some follow up papers \cite{Laine:2021ego,DeRocco:2021rzv,Ji:2021mvg}.
A recent quantum gravity conjecture suggests that trans-Planckian modes are not allowed to cross the horizon during inflation (TCC) which closely constrains the energy scale of inflation \cite{Bedroya:2019snp,Bedroya:2019tba}
Through a direct observational analysis \cite{Bastero-Gil:2017wwl,Lewis:2002ah}, we demonstrate that the minimal warm inflation is compatible with TCC and observations.  

\section{Minimal warm inflation} \label{sec:Cold}
In order to present minimal warm inflation, a direct coupling term is introduced between the axion (inflaton) and the gauge field in the Lagrangian \cite{Berghaus:2019whh}.
\begin{eqnarray}\label{Int}
\mathcal{L}_{int}=\frac{g_{YM}^2}{64\pi^2}\frac{\phi}{f}\tilde{G}_a^{\mu\nu}G_{\mu\nu}^a
\end{eqnarray}
$g_{YM}$ denotes the gauge coupling and $G_{\mu\nu}$ represents the field strength of any Yang-Mills group. 
Inflatons directly coupled to light gauge fields are able to reach thermal equilibrium suddenly. The process may begin with a small perturbation or even quantum fluctuations \cite{Berghaus:2019whh}.  
The effects of this interaction (\ref{Int}) have a significant impact on the inflaton (axion) field in the thermal bath, background, and perturbation.
The warm inflation scenario can be described by two coupled field equations. 
\begin{eqnarray}\label{EOM1}
\ddot{\phi}+(3H+\Upsilon(\phi,T))\dot{\phi}+V_{,\phi}=0\\
\nonumber
\dot{\rho}_{\gamma}+4H\rho_{\gamma}=\Upsilon\dot{\phi}^2\\
\nonumber
H^2=\frac{1}{3M_p^2}(\rho_{\phi}+\rho_{\gamma})~~~~~~\rho_{\gamma}=C_R T^4
\end{eqnarray} 
at the background level.
A direct interaction (\ref{Int}) yields the coupling term.  
As fields evolve at the slow-roll limit,
\begin{eqnarray}
\ddot{\phi}&\ll& 3H\dot{\phi},\\
\nonumber
\dot{\rho}_{\gamma}&\ll& 4H\rho_{\gamma},
\end{eqnarray}
 coupled equations (\ref{EOM1}) are simplified to
\begin{eqnarray}\label{EOM2}
3H(1+Q)\dot{\phi}\simeq -V_{,\phi}\\
\nonumber
4\rho_{\gamma}\simeq 3Q\dot{\phi}^2\\
\nonumber
H^2=\frac{V}{3M^2_p•}~~~~~~Q=\frac{\Upsilon}{3H}
\end{eqnarray} 
According to the minimal warm inflation model, dissipation coefficient  $\Upsilon$, which is originally derived via interaction term (\ref{Int}) in the Lagrangian, is proportional to the $T^3$.
We will study the hybrid form of the potential in detail \cite{Berghaus:2019whh}
\begin{eqnarray}\label{Hybrid}
V(\phi)=V_0+\frac{1}{2}m^2\phi^2
\end{eqnarray}
Using field evolution (\ref{EOM2}) in the slow-roll limit we can determine the evolution of dissipation parameter $Q$: \begin{eqnarray}\label{dissipation}
\frac{d \ln Q}{d N}=\frac{10\epsilon_V-\eta_V}{7Q}\\
\nonumber
\eta_V=M_p^2\frac{V_{,\phi\phi}}{V}~~~~~~~\epsilon_V=\frac{M_p^2}{2}(\frac{V'}{V})^2
\end{eqnarray}
which is a crucial relation in our study. 
For potential (\ref{Hybrid}), using Eq(\ref{dissipation}), we can calculate the dissipation parameter $Q$ based on the number of e-foldings $N$:
\begin{eqnarray}\label{Diss}
Q=\frac{m^2 M_p^2}{V_0}(\frac{6}{7}N+1)
\end{eqnarray} 
The power-spectrum of the high dissipative minimal warm inflation  model in terms of dissipation parameter $Q$ \cite{Berghaus:2019whh}
\begin{eqnarray}
\Delta_R=\frac{\sqrt{3}}{4\pi^{\frac{3}{2}}}\frac{H^3 T}{\dot{\phi}^2}(\frac{Q}{Q_3})^9 Q^{\frac{1}{2}}~~~~~~~~~\frac{T}{\dot{\phi}^2}=\frac{C_T}{4C_R H}
\end{eqnarray}
can be combined with relation (\ref{Diss}) and a relation for the number of e-folding $N$ based on wavenumber $\kappa$ \cite{Bastero-Gil:2017wwl}:
\begin{eqnarray}
N(k)=56.02-\ln \left(\frac{k}{k_0} \right)+\ln \left(\frac{V_k^{\frac{1}{2}}}{V_{\rm end}^{\frac{1}{2}}} \right)+\ln \left( \frac{V_{\rm end}^{\frac{1}{4}}}{10^{16}  \rm{GeV}} \right)
\end{eqnarray}
Therefore, we can write $ \Delta_R (\frac {k}{k_0}) $ of the minimal warm inflation model in the following way:
\begin{eqnarray} \label{eq:power-spectrum1}
\Delta_R \left(\frac{k}{k_0} \right)=\tilde{V}_0^{-8.5}\tilde{C}_T \left[75.5-0.86\ln \left(\frac{k}{k_0} \right)-0.21 \ln\tilde{V_0} \right]^{9.5}
\end{eqnarray}
where $\tilde{V}_0=(\frac{V_0}{m^2 M_p^2})^{-1}$, $\tilde{C}_T=2.2\times 10^{-5}C_T m^2$ are dimensionless parameters which are supposed to be constrained by observational data (see section (\ref{sec:observ})).
We will discuss what is known as trans-Planckian censorship conjecture (TCC) and how it constrains the field theories used to explain the inflation epoch. This would be interesting to show whether MWI is compatible with observational data and TCC criteria. 
\section{TCC and warm inflation} \label{sec:Warm}
\begin{figure}[h]
	\includegraphics[width=\columnwidth]{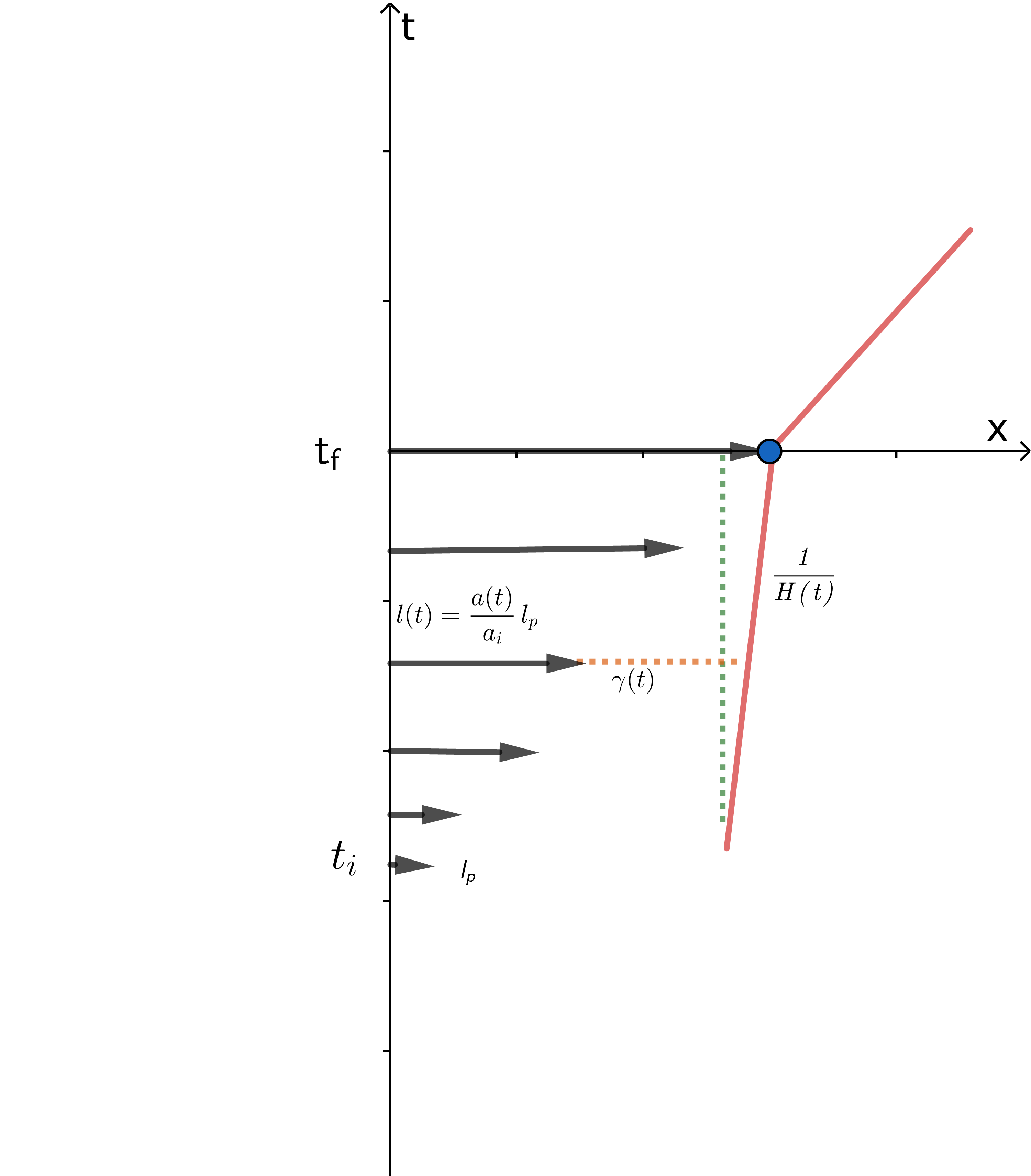}
	\caption{During an accelerated epoch all scales stretch. They can cross the horizon. Sub-plankean scale need more time for horizon crossing. To meet the TCC criteria (\ref{TCC}), the period of accelerated epoch need to be short enough. } 
	\label{index1}
\end{figure}

If we accept string theory as a consistent theory of quantum gravity, we can introduce a huge landscape of vacua which can be used in low energy IR limit. Based on swampland program some of the vacua that seem self-consistent and useful in IR limit are belong to the swampland. In this program there are some reasonable top-down constraints, in the limit of conjecture not theorem, on the filed excursion and the energy scale of the field theory that used in particle physics and cosmology \cite{Bedroya:2019snp, Bedroya:2019tba}. In the present work we focus on TCC which has a tight constraints on the energy scale of inflation. We will show the minimal warm inflation is compatible with criteria of TCC as well as observation in a wide range of parameter phase space. Sub-Planckian modes have to be hidden by Hubble horizon during inflation \ref{index1}:
\begin{eqnarray}
\label{TCC}
\frac{a_f}{a_i}.l_{p}<\frac{1}{H(t)}
\end{eqnarray}
which restricts the duration of inflation. If we assume constant Hubble horizon or quasi-de Siter inflation
\begin{eqnarray}\label{Constand Hubble}
H(t)\simeq H_f
\end{eqnarray}
we can find an upper limit for the energy density of inflation \cite{Bedroya:2019tba} 
\begin{eqnarray}\label{energy scale}
V^{\frac{1}{4}}<6\times 10^{8} \rm{GeV}
\end{eqnarray} 
and tensor to scale ratio parameter
\begin{eqnarray}
r<10^{-31}
\end{eqnarray}
which are nearly forbid the slow-roll inflation we need to resolve the shortcoming of HBB. But it would be relaxed if we consider
 warm inflation where the evolution of Hubble parameter during slow-roll epoch has an important role (\ref{index1}) on the theoretical perturbation parameters. In this case the tensor-to scalar ratio upper bound is more accessible \cite{Kamali:2019gzr}
\begin{eqnarray}
r<10^{-10}
\end{eqnarray} 
But swampland conditions would not be relaxed in the context of warm inflation \cite{Brandenberger:2020oav}.  
By a direct data analysis we try to find a parameter phase space of minimal warm inflation compatible with observation as well as TCC constrains.  
\section{Methodology and Observational Data} \label{sec:observ}
In this section, we describe the data and the analysis methods used here. We used most recent publicly available measurement of the CMB, released by \emph{Planck} team as \emph{Planck} 2018 data. For this purpose we used the Cosmic Microwave Background (CMB) temperature power spectrum (TT), the polarization power spectrum (EE), the cross-correlation power spectrum of temperature and polarization (TE), and the \emph{Planck} low-$\ell$ polarization likelihood (lowE) which we show by \emph{Planck} TT+Pol in figures and tables. ~\cite{Planck:2018vyg, Planck:2019nip}.\\
The primordial power spectrum of scalar perturbations  (Eq.~\ref{eq:power-spectrum1})  can be expressed as
\begin{equation}
\Delta_R \left(\frac{k}{k_0} \right)= \tilde{V}_1 \left[75.5-0.86\ln \left(\frac{k}{k_0} \right)-0.21 \ln\tilde{V_0} \right]^{9.5}
\end{equation}
which we redefined two parameters for convenience of calculation as follows
\begin{eqnarray}
\tilde{V}_0 = 10^7 \tilde{V}_0  \nonumber \\
10^{27} \frac{\tilde{C}_T}{\tilde{V}_0^{8.5}}= \tilde{V}_1
\end{eqnarray}
We consider a parameter space with six independent free parameters in our analysis:
$$
\mathbf{P} = \left\lbrace \Omega_b h^2, \Omega_c h^2 , 100 \theta_{\rm MC}, \tau, \tilde{V}_0 , \tilde{V}_1 \right\rbrace
$$
with flat uniform priors on the parameters reported in Table~\ref{tabpriors}. In our analysis, we assume curvature $\Omega_K$ is zero, and the Dark Energy equation of state is $w=-1$. We set the number of neutrino species $N_\nu=3.046$, and neutrino mass $m_\nu = 0.06 \rm{eV}$.\\
In our cosmological analysis, the allowed regions are calculated numerically with modified verision of the \textsc{CosmoMC} Markov Chain Monte Carlo (MCMC) sampler ~\cite{Lewis:2002ah} and the CAMB Boltzmann code. We apply Metropolis-Hasting sampling to 4 chains running in parallel, and use a convergence criterion that obeys $R-1 < 0.01$, where the Gelman-Rubin $R$-statistics ~\cite{Gelman92} is the variance of chain means divided by the mean of chain variances.\\
\begin{table}[tb]
\begin{center}
\begin{tabular}{lll}
\hline
\hline
Parameter & Symbol & Prior\\
\hline
Cold dark matter density & $\Omega_{\mathrm c}h^2$ & $[0.001, 0.99]$\\
Baryon density & $\Omega_{\mathrm b}h^2$ & $[0.005, 0.1]$\\
Warm inflation parameter & $10^7 \tilde{V}_0$ & $[0.1, 10.0]$\\
Warm inflation parameter & $ \tilde{V}_1$ & $[0.1, 10.0]$\\
Angular scale at decoupling & $100 \Theta_{\rm {MC}}$ & $ [0.5, 10.0]$\\
Optical depth & $\tau$ & $[0.01, 0.8]$ \\
Pivot scale $[{\rm{Mpc}}^{-1}]$ & $k_{\rm pivot}$ & 0.05 \\
\hline
\end{tabular}
\caption{Flat priors on the cosmological parameters varied in this paper.} \label{tabpriors}
\end{center}
\end{table}

\begin{table}[tb]
\begin{center}
\begin{tabular}{ c ||c  }
\hline
\hline
Parameters & \emph{Planck} \rm{TT}+\rm{Pol}\\
\hline
$\Omega_b h^2$ & $0.02124\pm 0.00012$ \\
$\Omega_c h^2$ & $0.1387\pm 0.0010$ \\
$10^7 \tilde{V}_0$ & $< 0.626$ \\
$\tilde{V}_1$ & $4.481^{+0.095}_{-0.20}$ \\
$100\theta_{MC}$ & $ 1.03888\pm 0.00029$ \\
$\chi^2_{\rm CMB}$ & $1359.4\,({\nu\rm{:}\,11.8})$ \\
\hline
\hline
\end{tabular}
\caption{$68\%$ CL constraints for the Warm inflation model, for \emph{Planck} 2018 temperature and polarization data.}
\label{table_results}
\end{center}
\end{table}

\begin{figure}[h]
\includegraphics[width=\columnwidth]{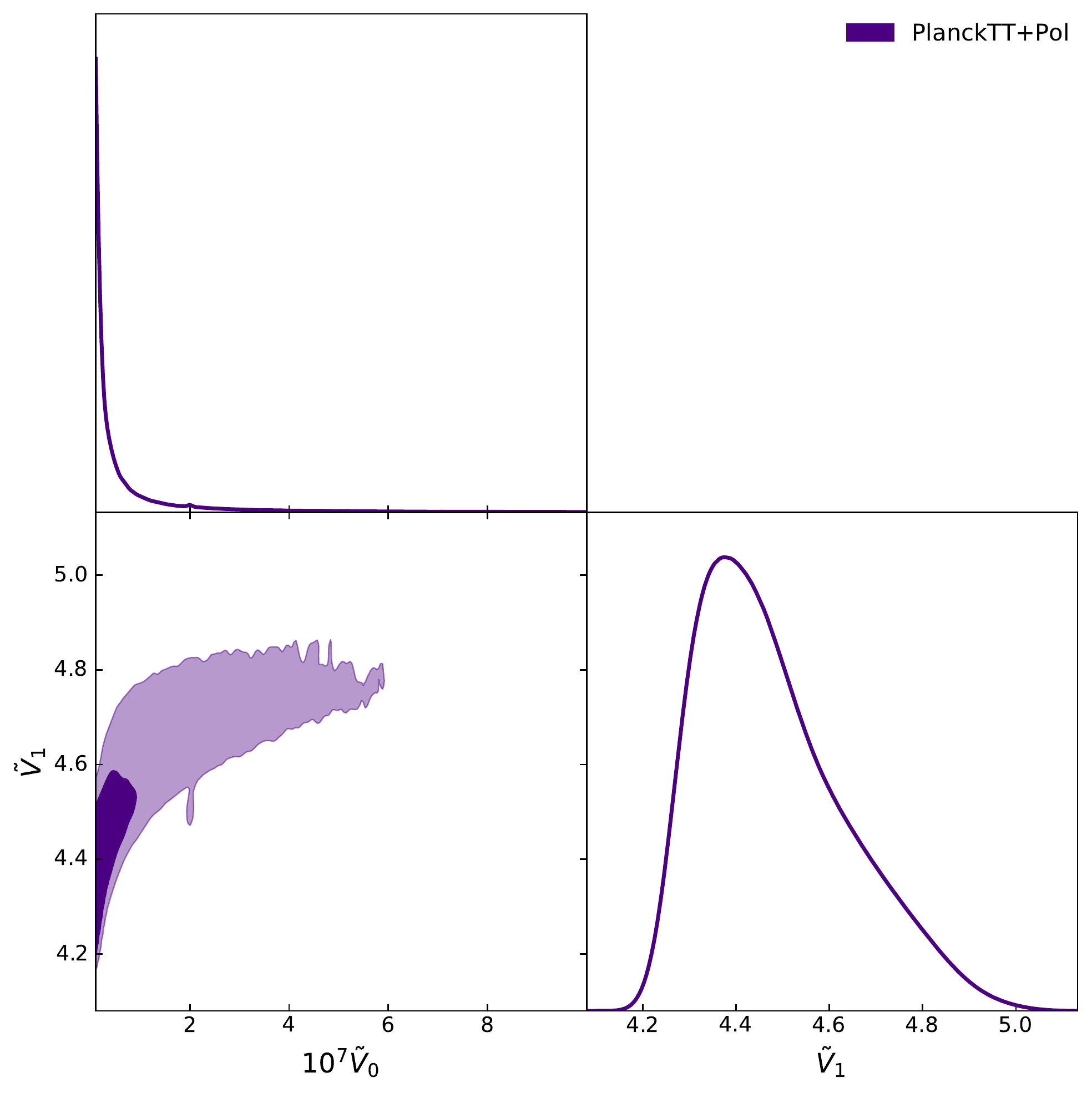}
\caption{Triangular plot for minimal warm inflation, representing contour plot and likelihoods of parameters of model $\tilde{V}_0$ and $\tilde{V}_1$. } \label{fig:tri-v0-v1}
\end{figure}

\begin{figure}[h]
\includegraphics[width=\columnwidth]{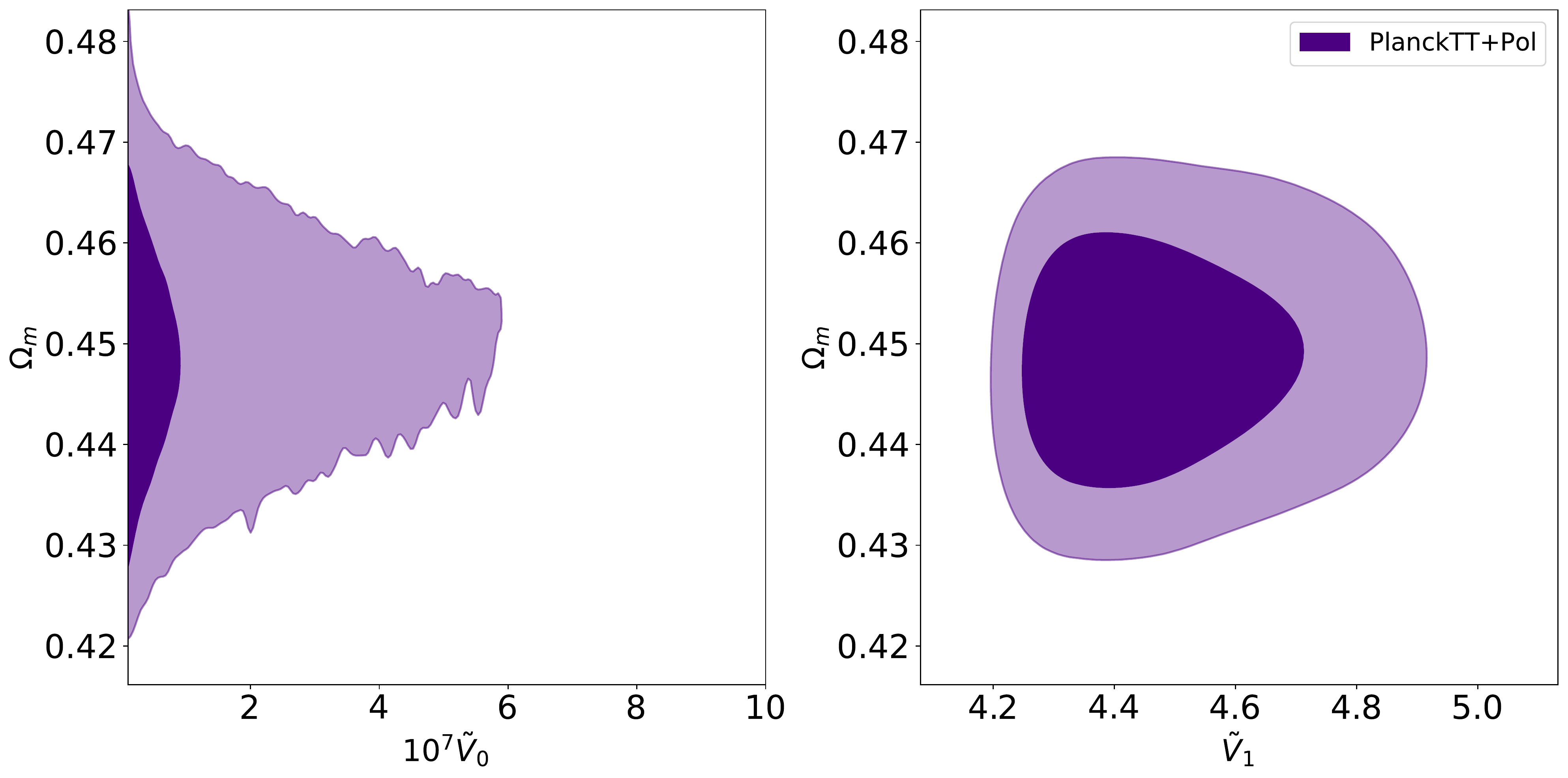}
\caption{Two-dimensional contour plots for minimal warm inflation, investigating the two parameters of model $\tilde{V}_0$ and $\tilde{V}_1$ vs. $\Omega_m$ are plotted. } \label{fig:2D_v0_v1_om}
\end{figure}

\begin{figure}[h]
\includegraphics[width=\columnwidth]{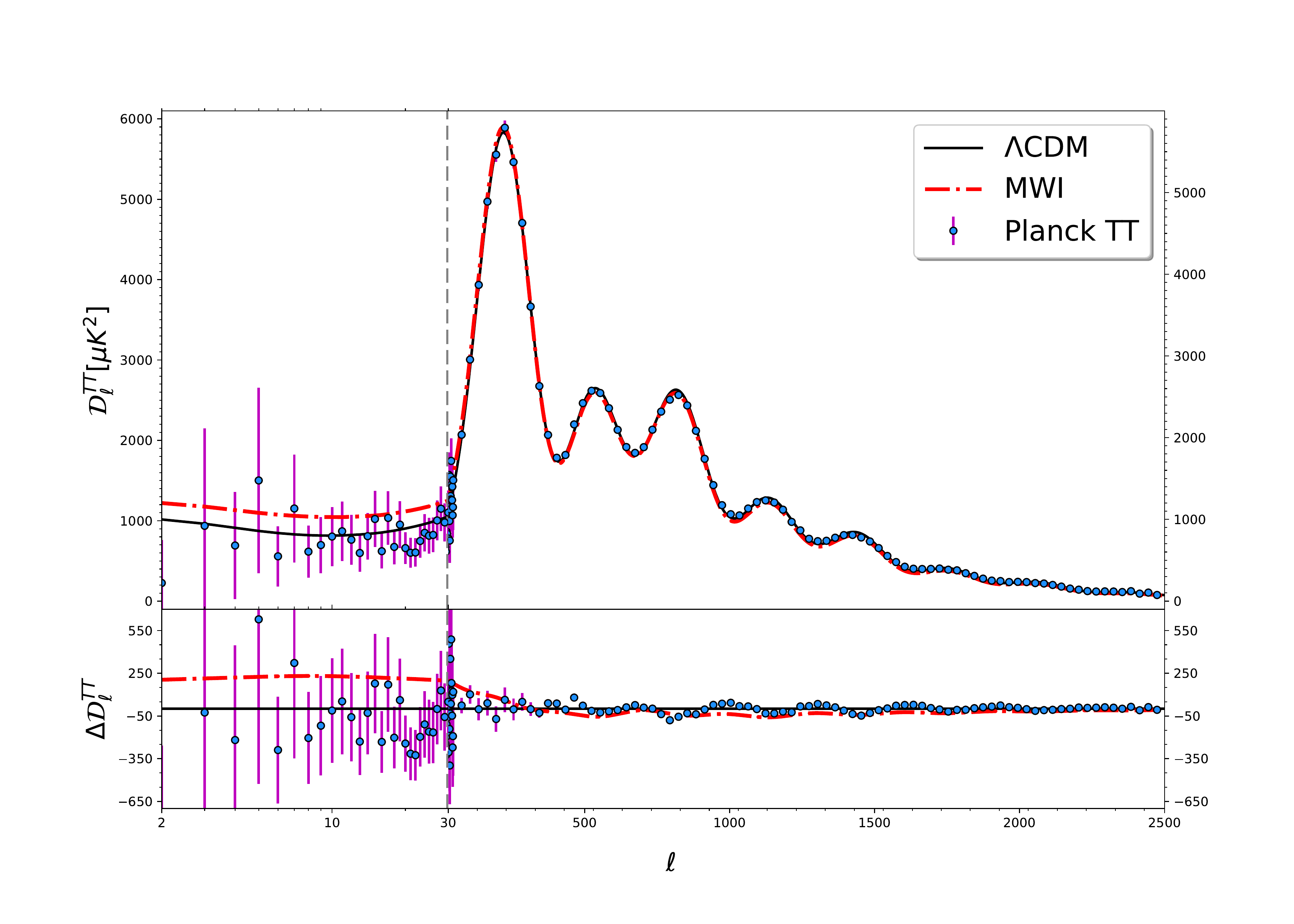}
\caption{The power spectrum of temperature anisotropies for minimal warm inflation model in comparison with standard $\Lambda$CDM model and \emph{Planck} 2018 power spectrum is plotted. Below panel shows the difference of MWI power spectrum and \emph{Planck} data. } \label{fig:power-cl}
\end{figure}

\section{Discussion and Conclusions} \label{sec:Conclusions}
In this paper we have tried to constrain  MWI model by using a direct method of data analysis. The Scalar power-spectrum of warm inflation as a function of wave number is presented by a non-standard model dependent format. It would be interesting to constrain the free parameters of the model directly. 
Best-fit values and confidence intervals for parameters of MWI model summarized in Table. \ref{table_results}. As you can see by implementation of just \emph{Planck} data we can put an upper bound on parameter $\tilde{V}_0 < 6.26 \times 10^{-8}$ which is consistent with TCC. Same result has shown in Fig. \ref{fig:tri-v0-v1}. In Fig. \ref{fig:2D_v0_v1_om} we see that matter density $\Omega_m$ shows nearly zero correlation with model parameters $\tilde{V}_0$ and $\tilde{V}_1$. Also we find that our model gives higher matter density than standard inflation. \\
The result we  have found for energy scales
\begin{eqnarray}
4 \times 10^{7}<V_0^{\frac{1}{4}}<5\times 10^{8} \rm{GeV}
\end{eqnarray}
of the model is surprisingly compatible with TCC (\ref{energy scale}) 
It is the main result of our discussion. 
   


\end{document}